\documentclass[...]{aa501}
\usepackage{graphicx}

\begin{document}

\title{A test for the search for life on extrasolar planets}
\subtitle{Looking for the terrestrial vegetation signature in the Earthshine
spectrum}


   \author{L. Arnold\inst{1},
          S. Gillet\inst{2}, O. Lardi\`ere\inst{2},
           P. Riaud\inst{2,3} and J. Schneider\inst{3}
          }

   \offprints{L. Arnold}

   \institute{Observatoire de Haute-Provence (OHP) CNRS 04870
Saint-Michel-l'Observatoire, France\\
             \email{arnold@obs-hp.fr}
         \and
             Laboratoire d'Interf\'erom\'etrie Stellaire et Exoplan\'etaire
(LISE) CNRS 04870 Saint-Michel-l'Observatoire, France\\
             \email{sgohp@obs-hp.fr, lardiere@obs-hp.fr, riaud@obs-hp.fr}
         \and
             Observatoire de Paris-Meudon, 92195, Meudon Cedex, France\\
             \email{Jean.Schneider@obspm.fr, Pierre.Riaud@obspm.fr}
             }

   \date{Received 24 December 2001; accepted 6 May 2002}

\abstract{ We report spectroscopic observations ($400 - 800\ nm$,
$R\approx100$) of Earthshine in June, July and October 2001 from
which normalised Earth albedo spectra have been derived. The
resulting spectra clearly show the blue colour of the Earth due to
Rayleigh diffusion in its atmosphere. They also show the
signatures of oxygen, ozone and water vapour. We tried to extract
from these spectra the signature of Earth vegetation. A variable
signal (4 to $10 \pm3\%$) around $700\ nm$ has been measured in
the Earth albedo. It is interpreted as being due to the vegetation
red edge, expected to be between 2 to $10\%$ of the Earth albedo
at $700\ nm$, depending on models. We discuss the primary goal of
the present observations: their application to the detection of
vegetation-like biosignatures on extrasolar planets.
\keywords{astrobiology -- stars: planetary systems} }
\authorrunning{Arnold {\it et al.}}
\titlerunning{The vegetation signature in the Earthshine spectrum}
\maketitle
%

\section{Introduction}

The search for life on extrasolar planets has become a reasonable goal
since the discovery of Earth-mass planets around a pulsar
(\cite{wolszczan_et_al92})
and Jupiter-mass planets around main-sequence stars (\cite{udry_et_al01}).
Although the detection of Earth-mass planets is not foreseen before space
missions  (like COROT scheduled for 2004, \cite{schneider_et_al98}), it is likely that
a significant  proportion of main sequence stars have Earth-like companions in their
habitable zone.  An important question is what type of biosignatures will unveil the possible
presence of life on these planets.

Spectral signatures can be of two kinds. A first type consists of
biological activity by-products, such as oxygen and its by-product
ozone, in association with water vapour, methane and carbon
dioxide (\cite{lovelock75}, \cite{owen80}, \cite{angel_et_al86}).
These biogenic molecules present attractive narrow molecular
bands. This led in 1993 to the Darwin ESA project
(\cite{leger_et_al96}), followed by a similar NASA project,
Terrestrial Planet Finder (TPF, \cite{angel_97},
\cite{beichman99}). But oxygen is not a universal by-product of
biological activity as demonstrated by the existence of anoxygenic
photosynthetic bacteria (\cite{blankenship_et_al95}).

\begin{figure}
   \centering
   \includegraphics[width=8.75cm]{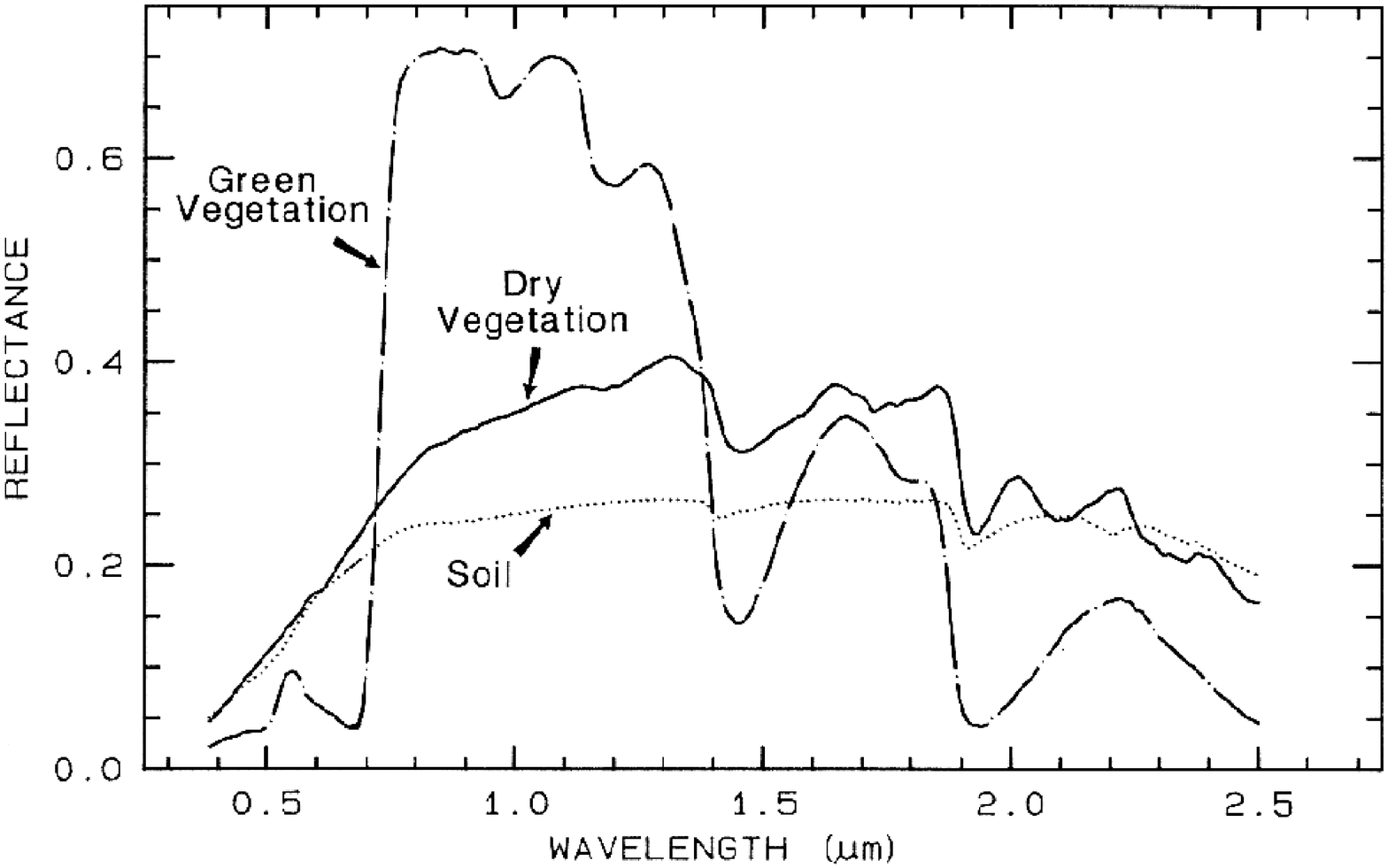}
   \caption{Reflectance spectra of photosynthetic (green) vegetation,
   non-photosynthetic (dry) and a soil (from \cite{clark99}).
   The so-called vegetation red edge (VRE) is the green vegetation
   reflectance strong variation from $\approx5\%$ at $670\ nm$
   to $\approx70\%$ at $800\ nm$.}
\label{vegetation}
\end{figure}

A second type of biosignature is provided by signs of stellar
light transformation into biochemical energy, such as the planet
surface colour from vegetation, whatever the bio-chemical details
(\cite{labeyrie99}). This must translates into the planet
reflection spectrum by some characteristic spectral features. This
signature is necessarily a more robust biomarker than any biogenic
gas such as oxygen, since it is a general feature of any
photosynthetic activity (here leaving aside chemotrophic
biological activity). Unfortunately, it is often not as sharp as
single molecular bands: although it is rather sharp for
terrestrial vegetation at $\approx700\ nm$ (\cite{clark99},
\cite{coliolo_et_al00}, see Fig.\ref{vegetation}), its wavelength
structure can vary significantly among bacteria species and plants
(\cite{blankenship_et_al95}).

Before initiating a search for extrasolar vegetation, it is useful
to test if terrestrial vegetation can be detected remotely. This
seems possible as long as Earth is observed with a significant
spatial resolution (\cite{sagan_et_al93}), but is it still the
case if Earth is observed as a single dot? A way to observe the
Earth as a whole is to observe the Earthshine with the Moon acting
like a remote diffuse reflector illuminated by our planet. It has
been proposed for some time (\cite{arci12}) to look for the
vegetation colour in the Earthshine to use it as a reference for
the search of chlorophyll on other planets, but up to now,
Earthshine observations apparently did not have sufficient
spectral resolution for that purpose (\cite{tikhoff14},
\cite{danjon28}, \cite{goode_et_al01}). We present in Section
\ref{section_EA} normalised Earth albedo spectra showing several
atmospheric signatures. We show in Section \ref{section_SR} how
the vegetation signature around $700\ nm$ can be extracted from
these spectra.


\section{Observations and data reduction} \label{section_obs}
After a first test made in 1999 with the FEROS spectrograph
($R=48000$) on the La Silla ESO $1.5\ m$ telescope, we have built
a dedicated spectrograph mounted on the $80\ cm$ telescope at the
Observatoire de Haute-Provence (Table \ref{tel_spec_char}).

We have observed the Moon at ascending and descending phases from April to
October 2001. Data collected in June, July and October (Table \ref{journal}) have been of
sufficient quality to derive the results described in this article.

Our observation procedure is the following. The long spectrograph slit allows
us to record simultaneously the Earthshine and sky background spectra ($\approx 80$
CCD lines for each). This single
exposure is bracketed by two spectra of the Moonlight. Each of the latter is
the mean of 10 spectra (totalling $20\ \arcmin$) taken in different regions to
smooth the Moon albedo spatial variations. A series of
flat fields (tungsten lamp) is recorded just after the previous cycle.

Before getting a final spectrum from binning of CCD lines, a
sub-pixel alignment of CCD lines is done to correct the residual
angle between the pixel rows and the dispersion direction: Each
image is oversampled by a factor of 8 in the dispersion direction.
Each line $i$ is then translated to maximize the cross-correlation
function from line $i$ with line $1$. After the line binning is
done, the spectrum is resampled with $1\ nm/pixel$ for
convenience.

\begin{table}
      \caption[]{Telescope and spectrograph characteristics.}
         \label{tel_spec_char}
     $$
         \begin{array}{p{0.5\linewidth}l}
            \hline
            \noalign{\smallskip}
            Parameters   &  $Values$  \\
            \noalign{\smallskip}
            \hline
            \noalign{\smallskip}
            Telescope diameter, f/ratio     & 80\ cm, 16.5   \\
            Slit length  (unvignetted)      & 6\ arcmin\ (2\ arcmin)             \\
            Slit width                      & 1.6\ $to$\ 7.8\ arcsec  \\
            Spatial sampling                & 1.4\ arcsec/pixel       \\
            Transmission grating            & 100\ lines/mm           \\
            Spectral resolution $\lambda/\Delta\lambda$   & \approx 50\ $to$\ 240\ $at$\ \lambda=700\ nm\\
            Max. spectral range             & 400\ $to$\ 900\ nm      \\
            Spectral sampling               & 2.6\ nm/ pixel                \\
            CCD Kodak ship          & $non-ABG$\ $KAF-0401E$  \\
            \noalign{\smallskip}
            \hline
         \end{array}
      $$
\end{table}

\begin{table}
      \caption[]{Earthshine observations journal.}
         \label{journal}
     $$
         \begin{array}{p{.26\linewidth}llll}
            \hline
            \noalign{\smallskip}
            Date           & $Hour$           & $Exposure$               & $Resolution$      &   $Signal$ \\
            (yyyy/mm/dd)   & $(h\ min\ UT)$   & $time\ (s)$^{\mathrm{a}} & \lambda/\Delta\lambda & $to Noise$^{\mathrm{b}}\\
            \noalign{\smallskip}
            \hline
            \noalign{\smallskip}
            2001/06/17 & 02\ 37\ $to$\ 03\ 08  &   \ \ \ \  480 & \ \ 120  & 190 \\
            2001/06/18 & 02\ 49\ $to$\ 03\ 07  &   \ \ \ \  360 & \ \ 120  &  130\\
            2001/06/24 & 20\ 20\ $to$\ 21\ 03  &   \ \ \ \  600 & \ \ 120  &  210\\
            2001/06/26 & 20\ 41\ $to$\ 21\ 45  &   \ \ \ \  900 & \ \ 120  &  150\\
            2001/07/23 & 20\ 04\ $to$\ 20\ 37  &   \ \ \ \  240 & \ \ 50   &  170\\
            2001/07/24 & 19\ 54\ $to$\ 20\ 56  &   \ \ \ \  1080 & \ \ 50  &  390\\
            2001/07/25 & 19\ 57\ $to$\ 21\ 24  &   \ \ \ \  1440 & \ \ 50  & 250 \\
            2001/10/13 & 02\ 16\ $to$\ 05\ 00  &   \ \ \ \  2640 & \ \ 240 & 240 \\
            2001/10/14 & 03\ 35\ $to$\ 04\ 04  &   \ \ \ \  480 & \ \ 240 & 100 \\
            2001/10/19 & 17\ 37\ $to$\ 17\ 54  &   \ \ \ \  240 & \ \ 240 & 40 \\
            2001/10/21 & 17\ 39\ $to$\ 19\ 25  &   \ \ \ \  1680 & \ \ 240 & 150 \\
            \noalign{\smallskip}
            \hline
         \end{array}
      $$
\begin{list}{}{}
\item[$^{\mathrm{a}}$] Cumulative exposure time from several $120s$, $180s$
or $240s$ single exposures.

\item[$^{\mathrm{b}}$] Signal to noise ratio at $\lambda=650\ nm$ for the
cumulative spectrum obtained after single images addition and
lines binning.
\end{list}
\end{table}

\section{Results}
\subsection{Earth albedo $EA(\lambda)$}
\label{section_EA} Let us define the following spectra: we call
the Sun as seen from outside the Earth atmosphere $S(\lambda)$,
Earth atmosphere transmittance $AT(\lambda)$, Moonlight
$MS(\lambda)$, Earthshine $ES(\lambda)$, Moon albedo
$MA(\lambda)$, and Earth albedo $EA(\lambda)$. We have
\begin{eqnarray}
MS(\lambda) = S(\lambda)\times MA(\lambda) \times AT(\lambda) \times g_1,
\label{MS}
\end{eqnarray}
\begin{eqnarray}
ES(\lambda) = S(\lambda)\times EA(\lambda)\times MA(\lambda) \times
 AT(\lambda) \times g_2.
\label{ES}
\end{eqnarray}
The Earth albedo is simply given by Eq.\ref{ES}/Eq.\ref{MS}, i.e.
\begin{eqnarray}
EA(\lambda) = {ES(\lambda) \times g_1 \over MS(\lambda) \times g_2}.
\label{EA}
\end{eqnarray}
Simplifying by $AT(\lambda)$ means that $ES(\lambda)$ and
$MS(\lambda)$ should be ideally recorded simultaneously to avoid
significant airmass variation and thus Rayleigh scattering bias.
The mean of the two $MS$ spectra bracketing $ES(\lambda)$ is thus
used to compute $EA(\lambda)$. The $g_i$ terms are geometric
factors related to the Sun, Earth and Moon positions. For
simplicity, we set $g_1$ and $g_2$ equal to 1, equivalent to a
spectrum normalization.

\begin{figure}
   \centering
   \includegraphics[width=8.75cm]{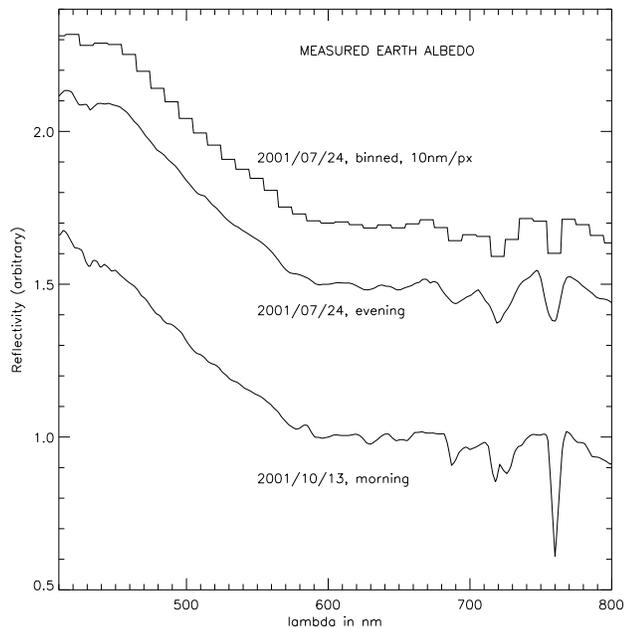}
   \caption{Examples of measured Earth albedo spectra.  Both spectra are
    normalized to 1 at $600\ nm$, but the July spectrum is shifted upwards
    by 0.5 for clarity. The spectral resolution was
    $\approx50$ in July, and $\approx240$ in October.
    The July spectrum has been binned to $10\ nm/px$ to mimic the low
    resolution that might be used for the first extrasolar planet spectrum.}
    \label{figure_EA}
\end{figure}

Fig.\ref{figure_EA} shows that the Earth should be seen as a blue
object from space. This blue colour has been known for a long
time, and has been confirmed by the Apollo astronauts
(\cite{kelley88}). Tikhoff (1912) had already discovered the blue
colour of Earthshine and interpreted it as being due to the
Rayleigh scattering in the atmosphere. This point will be
discussed in more details in Sect.\ref{section_SR}.

The $H_2O$ bands around $690$ and $720\ nm$, and $O_2$ narrower
band at $760\ nm$ are clearly visible with a resolution of
$R\approx50$. The slope variation occurring at $\approx600\ nm$ is
partially the signature of the deepest zone of the broad ozone
absorption band (Chappuis band), going from $440$ to $760\ nm$.

\subsection{Earth surface reflectance $SR(\lambda)$}
\label{section_SR} To detect the vegetation signature at $700\
nm$, it is necessary to extract the Earth surface reflectance
$SR(\lambda)$, that contains the spectral information on
vegetation, from the atmosphere features contained in
$AT(\lambda)$. Said differently, it is necessary to remove the
atmospheric bands in this spectral region. Surface reflectance
$SR(\lambda)$ is usually presented in Earth remote sensing science
by a vector giving the directional properties of the scattered
light (\cite{liang99, brdf00}). But here, we adopt a simpler
scalar definition allowing us to write the albedo spectrum
$EA(\lambda)$ as the product
\begin{equation}
EA(\lambda) \approx SR(\lambda)\times AT^{\alpha=2}(\lambda)
\label{refatm2}
\end{equation}
meaning that photons are transmitted once through a one airmass
Earth atmosphere, are scattered by the Earth's surface, and then
are transmitted back through the atmosphere a second time, giving
a power of 2 on $AT(\lambda)$. Clearly $\alpha$ represents an
airmass, but its value of 2 is a rough approximation: all photons
do not cross twice an airmass of 1, depending on their impact
location on Earth, on how they are scattered in the Earth's
atmosphere versus their wavelength, and again how the
Sun-Earth-Moon triplet is configured (described by a
time-dependent vector $\vec{g_3}$). Moreover, photons can be
reflected by high-altitude clouds having a high albedo, thereby
crossing a thinner airmass before going back to space. The latter
proportion of photons is also time-dependent, thus implying that
$\alpha=\alpha(\vec{g_3},\lambda,t)$ is probably difficult to
estimate.

We obtained $AT(\lambda)$ by the ratio of two mean
spectra $MS(\lambda)$ taken at two different Moon elevations. This obviously
gives only a measure of the local atmosphere transmittance, whereas $AT(\lambda)$ in Eq.\ref{refatm2} represents
a mean spectrum for the illuminated Earth seen from the Moon.

\begin{figure}
   \centering
   \includegraphics[width=8.75cm]{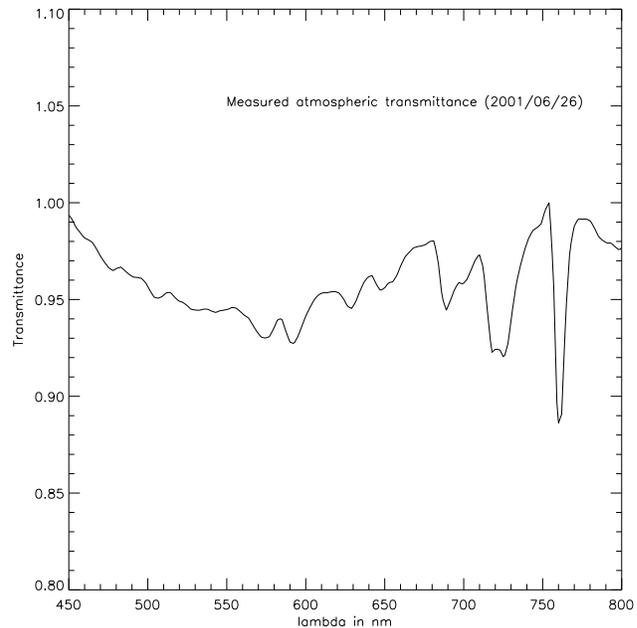}
      \caption{An example of a measured
1-airmass atmospheric transmittance, after Rayleigh scattering
correction. $H_2O$ and $O_2$ bands are obviously present, while
the broad ozone goes from 450 to $700\ nm$ with maximum absorption
at $\approx600\ nm$.} \label{figure_AT}
\end{figure}

\begin{figure}
   \centering
   \includegraphics[width=8.75cm]{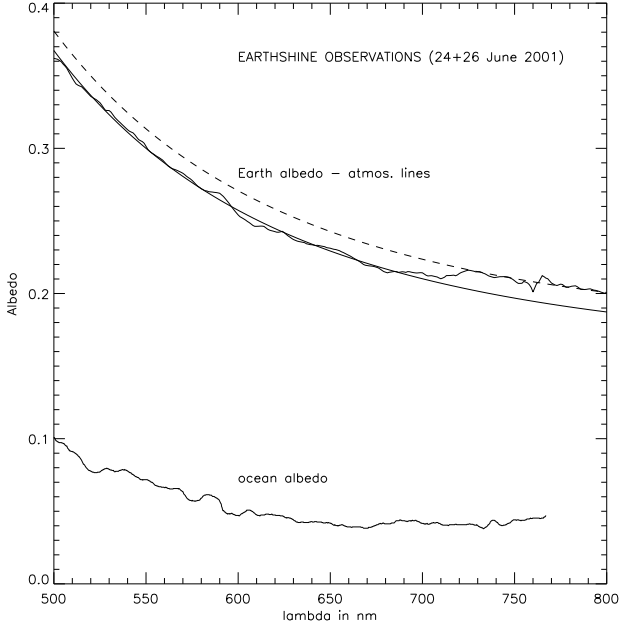}
      \caption{An example of Rayleigh correction: The graph shows
the 24+26 June spectrum $SR(\lambda)$ (above) after atmospheric
absorption correction (Eq.\ref{SR}), but still containing the
Rayleigh scattering signature. The spectrum is fitted with a
Rayleigh law adjusted over the [500;670nm] window. The fit is then
translated (dash) and adjusted to the [740;800nm] region of
$SR(\lambda)$ to show the VRE (here $VRE=7\%$). $SR(\lambda)$ is
normalised to 0.3 at $550\ nm$ (\cite{goode_et_al01}) to compare
to the ocean albedo (\cite{mclinden_et_al97}). The $SR(\lambda)$
higher slope in the blue is the signature of Rayleigh diffusion in
Earth's atmosphere. }
         \label{fig_rayleigh}
\end{figure}

Our measured $AT(\lambda)$ is corrected for Rayleigh scattering
and is normalized to 1 airmass (Fig.\ref{figure_AT}). Then
Eq.\ref{refatm2} gives
\begin{equation}
SR(\lambda) =  \frac{EA(\lambda)}
               {AT^\alpha(\lambda)}.  \label{SR}
\end{equation}
We adjusted $\alpha$ to remove the atmospheric bands in
$SR(\lambda)$ between 600 and $800\ nm$. We found $\alpha$ values
between 1 and 3. Most of the time, we treated the $O_2$, $O_3$ and
$H_2O$ bands separately with different $\alpha$ to obtain the best
correction.

 The surface reflectance we obtain in practice from
Eq.\ref{SR} is shown in Fig.\ref{fig_rayleigh}.  The albedo
spectrum of the ocean is $\approx0.1$ at $500\ nm$ and decreases
smoothly to $\approx0.05$ at $750\ nm$ (\cite{mclinden_et_al97},
Fig.\ref{fig_rayleigh}). But considering these values and the land
albedo from Fig.\ref{vegetation}, associated with a typical cloud
cover of 50\% for both ocean and land, it can be shown that the
higher Earth albedo in the blue in Fig.\ref{figure_EA} cannot be
explained by the higher albedo of the ocean in the blue, but
rather by a contribution of Rayleigh diffusion in the atmosphere.

\begin{figure}
   \centering
   \includegraphics[width=8.75cm]{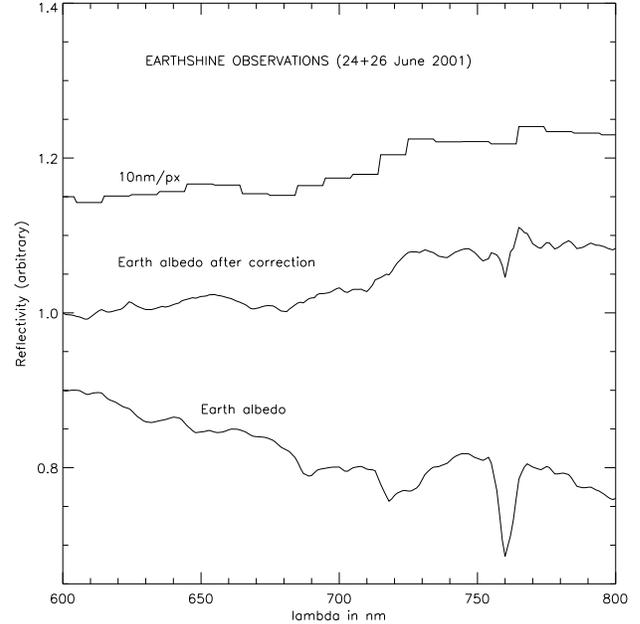}
      \caption{An example of data reduction sequence: The graph shows
the June albedo spectrum $EA(\lambda)$ (bottom). All atmospheric
absorption features are then corrected according to Eq.\ref{SR}
and  the spectrum is flattened with a Rayleigh law adjusted in the
[500;670nm] window (Fig.\ref{fig_rayleigh}). The result is shown
above with 1 and $10\ nm/px$ resolution. The measured red edge
around $700nm$ is $VRE=7\%$ (Eq.\ref{VRE}).}
         \label{fig_VRE}
\end{figure}

Therefore $SR(\lambda)$ does not represent the pure surface
reflectance, but includes uncorrected atmospheric scattering (for
simplicity, we nevertheless continue to name $SR(\lambda)$ the
result of Eq.\ref{SR}). The Fig.\ref{fig_rayleigh} shows
$SR(\lambda)$ fitted with the Rayleigh law $A+B/\lambda^4$
adjusted over the [500;670nm] window. The slope towards the blue
does not hide the relatively sharp vegetation signature, which
appears around $700\ nm$. $SR(\lambda)$ is then normalized to the
Rayleigh fit (Fig.\ref{fig_VRE}).

\begin{figure}
   \centering
   \includegraphics[width=8.75cm]{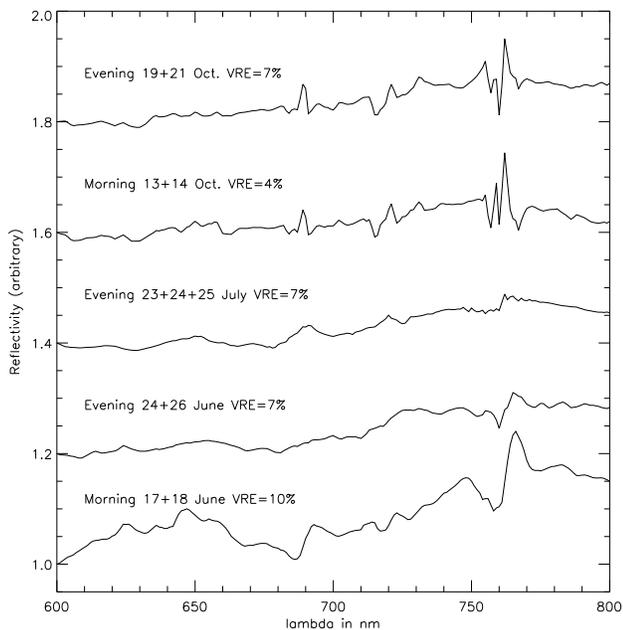}
      \caption{Collection of $SR(\lambda)$  spectra
normalized to 1 at $600\ nm$, but shifted upwards for clarity by
0.2, 0.4, 0.6 and 0.8, respectively. Note that only the [600;670]
and [740;$800\ nm$] windows are used to estimate the VRE.}
\label{fig_VRE_all}
\end{figure}

\begin{figure}
   \centering
   \includegraphics[width=8.75cm]{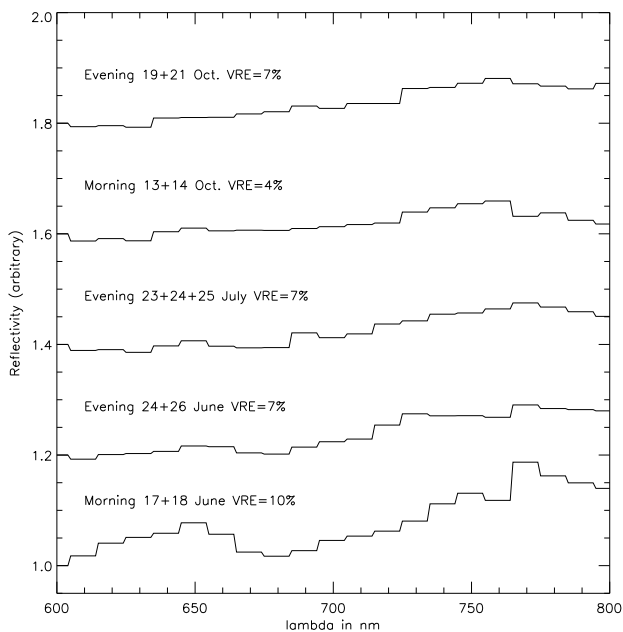}
      \caption{The same collection of $SR(\lambda)$ spectra as in Fig.\ref{fig_VRE_all} but
binned to $10\ nm/px$. } \label{fig_VRE_binned_all}
\end{figure}

To quantify the vegetation signature, we define the Vegetation Red Edge ($VRE$) as
\begin{equation}
VRE =  \frac{r_I-r_R}{r_R}  \label{VRE}
\end{equation}
where $r_R$ and $r_I$ are the mean reflectances in the [600;670nm] and
[740;800nm] windows in the spectrum after it has been flattened
with a Rayleigh law as explained above.
This $VRE$ definition giving the relative height of the step due to the
vegetation is close to the $NDVI$
(Normalized Difference Vegetation Index, \cite{rouse_et_al74, tucker79})
used in Earth satellite observation which considers
the difference, after atmospheric correction, between  the reflected fluxes in broad red
and infra-red bands, normalized to the sum of the fluxes in these bands,
\begin{equation}
NDVI =  \frac{f_I-f_R}{f_R+f_I}  \label{NDVI}.
\end{equation}
Flattened $SR(\lambda)$ spectra are shown in Fig.\ref{fig_VRE_all}
and \ref{fig_VRE_binned_all}. Eq.\ref{VRE} gives VRE values
ranging between 4 and 10\%.

\section{Discussion}
\label{discussion} First, we verified that the Moon albedo cancels
correctly in practice when we do the ratio
$EA(\lambda)=ES(\lambda)/MS(\lambda)$ (Eq.\ref{EA}). Since
$MS(\lambda)$ is the mean of 10 long-slit Moonlight spectra
totalling $20\arcmin$ while a single Earthshine $ES(\lambda)$
spectra is only $2\arcmin$, we computed the ratio of 2 spectra
($2\arcmin/20\arcmin$) of Moonlight $MS(\lambda)$ taken on
different Moon regions. We obtain a constant with $\sigma<0.5\%$
meaning that the Moon albedo cancels correctly in Eq.\ref{EA}.

We also verified that the second order spectrum pollution is negligible: it has been
measured to be $0\%$ at $760\ nm$, $<0.3\%$ at $780\ nm$ and $<0.5\%$ at $800\ nm$.

To test our VRE measurements, we also measured the VRE on spectra
of Vega and the sunlit Moon, for which we obviously should have
$VRE=0\%$. After standard bias, dark and flat corrections, the
data are calibrated with reference A0V and Sun spectra,
respectively, then flattened by normalization to a black body
curve. Vega spectra show a $VRE=-1\%,\ \sigma=2\%$. We also used
the standard Moon albedo spectrum of Apollo16 sample 62231
(\cite{pieters99}) to properly flatten the Moon data: We obtain
$VRE=0\%,\ \sigma=3\%$. Therefore, we conclude that {\it i)} the
VRE on these sources is measured to be $VRE=0\%$, with
$\sigma\approx3\%$, and {\it ii)} the VRE we measured in the Earth
albedo ranges between 4 and 10\%, with a similar error of
$\sigma\approx3\%$ (Table \ref{table_VRE}). In the June 17+18
spectrum, we probably have a larger error on the $VRE$ due to the
lack of bracketing from Earthshine with Moonlight spectra
(Fig.\ref{fig_VRE_all} and \ref{fig_VRE_binned_all}).

\begin{table}
      \caption[]{Results for the measured Vegetation Red Edge $VRE$.}
         \label{table_VRE}
     $$
         \begin{array}{lll}
            \hline
            \noalign{\smallskip}
            $Epoch$                 & $Viewed zone$         & VRE \\
            \hline
            $2001/06/17\&18 morning$    & $Eur./Asia$         &10 \pm 5\% \\
            $2001/06/24\&26 evening$    & $Amer./Pacific$     &7 \pm 3\%\\
            $2001/07/23\&24\&25 evening$    & $Amer./Pacific$ &7 \pm 3\%\\
            $2001/10/13\&14 morning$    & $Eur./Asia$         &4 \pm 3\%\\
            $2001/10/19\&21 evening$    & $Amer./Pacific$     &7 \pm 3\%\\
            \hline
         \end{array}
      $$
\end{table}

These results are in agreement with estimations from models:
\cite{desmarais_et_al01} predict a vegetation signature of ``$2\%$, maybe larger if a large forested
area is in view''. Preliminary estimates gave 5\% (\cite{schneider_00a}, \cite{schneider_00b}), while
our model described hereafter predicts $\approx10 \%$.

Only green lands observed by different Earth observing satellites
show this spectral feature around $700\ nm$. It is thus legitimate
to attribute our global VRE to the terrestrial vegetation. Let us
investigate this hypothesis a little further.

Table \ref{zone} shows which portion of the Earth is seen from the
Moon at the time the observations were done. An important
contribution to the Earth albedo comes from clouds. When clouds
cover land without vegetation or an ocean area, its albedo adds to
the general planet albedo, without suppressing the vegetation
contribution, and simply reduces the global VRE. When clouds cover
a forest region, the vegetation contribution to the VRE for that
Earth region is removed. The global VRE thus obviously depends not
only on the global cloud cover, but also on which regions are
covered by clouds during the observation. We have therefore taken
cloud cover images available from
http://wwwghcc.msfc.nasa.gov/GOES/globalir.html for each of our
observation dates and have estimated for each run of observation
the fraction of land and ocean covered by clouds (Table
\ref{cloud}).

One can  {\it a priori} estimate the $VRE$ from the albedo and
relative surfaces of ocean, land and cloud. Using Eq.\ref{VRE},
the $VRE$ can be written

\begin{eqnarray}
  VRE_{th} =  
\frac{A_l(r_I-r_R)_l CF_l}
{A_o(r_R)_o CF_o+A_l(r_R)_l CF_l + CC(r_R)_c}
  \label{VRE_th}
\end{eqnarray}
where $CF_{o, l}$ are the fractional cloud-free areas for ocean
and land (from deserts to green forests), respectively, and $A_{o,
l}$ are the corresponding areas in view from the Moon. $CC$ is the
fractional cloud cover, equal to $CC=(1-CF_o)A_o+(1-CF_l)A_l$.
Calculating a mean $VRE_{th}$ for Earth  with $A_o=0.7$ and
$A_l=0.3$ (standard Earth ocean/land ratio), $(r_I-r_R)_l = 0.15$
(based on $NDVI$ data), $(r_R)_o=0.05$, $(r_R)_l=0.25$ and
$(r_R)_c=0.5$ (albedos for ocean, land and cloud), we find, for a
mean cloud cover of $50\%$ for all regions, $VRE_{th}\approx7\%$.
We estimate the error on $VRE_{th}$ to be $\pm3\%$ (cloud cover
estimation, seasonal variation of the difference $(r_I-r_R)_l$,
vegetation mutual shadow effects). The small contribution of ocean
chlorophyll is also neglected here.

The resulting $VRE_{th}$ computed from the Earth phase seen from
the Moon (Table \ref{zone}) and cloud cover (Table \ref{cloud})
are given in Table \ref{table_VRE_obs_th}. The denominator of
Eq.\ref{VRE_th} represents an albedo, known to be 0.30 for the
Earth (\cite{goode_et_al01}), and its value here ranges between
0.30 and 0.34 depending on the date of observation. The $VRE_{th}$
values are in acceptable agreement with preliminary estimates of
5\% (\cite{schneider_00a}, \cite{schneider_00b}), but are higher
than our observations. This partially comes from terrestrial limb
and directionality of the vegetation reflectance effects. The
latter are complicated (\cite{privette_et_al95},
\cite{hapke_et_al96}), due for instance to mutual shadow effects
between trees.

\begin{table}
      \caption[]{Earth zone viewed from the Moon at the epochs of observations. }
         \label{zone}
     $$
         \begin{array}{llll}
            \hline
            \noalign{\smallskip}
            $Epoch$   &  $Viewed\ $ $zone$ & $Land$     & $Ocean$\\
                      &                    &  \%          & \% \\
            \noalign{\smallskip}
            \hline
            \noalign{\smallskip}
            $2001/06/17\&18$   & $Eur./Asia$              & 50 & 50\\
            $2001/06/24\&26$   & $Amer./Pacific$          & 40 & 60\\
            $2001/07/23\&24\&25$   & $Amer./Pacific$          & 40 & 60\\
            $2001/10/13\&14$   & $Eur./Asia$              & 50 & 50\\
            $2001/10/19\&21$   & $Amer./Pacific$          & 30 & 70\\
            \noalign{\smallskip}
            \hline
         \end{array}
      $$
\end{table}

\begin{table}
      \caption[]{Cloud cover estimated for the Earth zone viewed from the Moon at the epochs of observations.}
         \label{cloud}
     $$
         \begin{array}{lcc}
            \hline
            \noalign{\smallskip}
            $Epoch$                     & $Land$                & $Ocean$    \\
                                        & (1-CF_l) \%        & (1-CF_o) \% \\
            \hline
            $2001/06/17\&18 morning$      & 60                    &50\\
            $2001/06/24\&26 evening$      & 40                    &50 \\
            $2001/07/23\&24\&25 evening$      & 50                    &50\\
            $2001/10/13\&14 morning$      & 50                    &50\\
            $2001/10/19\&21 evening$      & 50                    &50\\
            \hline
         \end{array}
      $$
\end{table}

\begin{table}
      \caption[]{Comparison of the observed and calculated Vegetation Red
 Edge $VRE$.}
         \label{table_VRE_obs_th}
     $$
         \begin{array}{lccc}
            \hline
            \noalign{\smallskip}
            $Epoch$   & $Viewed zone$   & VRE_{obs}    & VRE_{th} \\
            \hline
            $2001/06/17\&18 morning$ & $Eur./Asia$         &10 \pm 5\%    & \ 9  \pm 3\%  \\
            $2001/06/24\&26 evening$ & $Amer./Pacific$     &7 \pm 3\%     & 12 \pm 3\%  \\
            $2001/07/23\&24\&25 evening$ & $Amer./Pacific$     &7 \pm 3\% & \ 10  \pm 3\%  \\
            $2001/10/13\&14 morning$ & $Eur./Asia$         &4 \pm 3\%     & 12 \pm 3\%  \\
            $2001/10/19\&21 evening$ & $Amer./Pacific$     &7 \pm 3\%     & \ 7  \pm 3\%  \\
            \hline
         \end{array}
      $$
\end{table}

\section{Conclusion}

Although it seems that the  Earth's vegetation signature might be
visible as a red edge at $700\ nm$, it is difficult to measure  in
the Earthshine for two  reasons. The first reason is related to
its variable amplitude, induced by a variable cloud cover and
Earth phase. The second reason is because it is hidden below
strong atmospheric bands which need to be removed to access the
surface reflectance including the vegetation signature.

For the Earth, our knowledge of different surface reflectivities
(deserts, ocean, ice etc) help us to assign  the VRE of the
Earthshine spectrum to terrestrial vegetation. For an exoplanet, a
VRE-like index might be as difficult to measure as for the Earth
due to variable cloud cover of the planet. Even if an extrasolar
planet would give a clear VRE-like spectral signal, its use as a
biosignature would raise some questions because:

1/ For several organisms (such as {\it Rhodopseudonomas},
\cite{blankenship_et_al95}) the ``red edge'' is not at $700\ nm$, but at $1100\ nm$.

2/ Some rocks, like schists, may have a similar spectral feature. For instance,
spectra of Mars show a similar spectral feature at $3.5 \mu$, which
were erroneously interpreted as vegetation due to their similarity
with lichen spectra (\cite{sinton57}).

We nevertheless believe that, associated with the presence of
water (and secondarily oxygen) and correlated with seasonal
variations, a vegetation-like spectral feature would provide more
insight than simply oxygen on the bio-processes possibly taking
place on the planet. But since water, and thus clouds and rain,
are essential for the growth of vegetation, extrasolar planets
with a very low cloud cover and a corresponding high vegetation
index are unlikely, more especially if the planet is seen pole-on,
with a bright white polar cover. On the other hand, an extrasolar
planet vegetation surface could be larger than on Earth (like
during periods in the paleozoic and mesozoic eras on Earth for
example).

One must also note that the measurement of an extrasolar planet
VRE will not suffer from the intrinsic difficulty of the same
measurement for the Earth through the Earthshine spectrum: The
extrasolar planet albedo will simply be given by the ratio of
spectra $planet/mother\ star$. But a model of the exoplanet
atmosphere is necessary to be able to remove the absorption bands
that may partially hide the vegetation. The detection of a VRE
index between 0 and 10\% requires a photometric precision better
than 3\%. Exposure time to achieve this precision with Darwin/TPF
on an Earth-like planet at $10\ pc$ with a spectral resolution of
25 is of the order of $100\ h$ based on recent simulations
(\cite{riaud_et_al02}).

Finally, the Earth albedo spectral variations study is of interest
for global Earth observation. It might provide data on climate
change, as broad-band measurements recently showed
(\cite{goode_et_al01}). We also think that the spectrum of
Earthshine might be used for example to monitor the global ozone
(with the Chappuis or Huggins bands).

During the submission of the paper, we have been informed of a similar work by \cite{woolf_et_al02}.

\begin{acknowledgements}
We are grateful to  P. Fran\c{c}ois who took test spectra of the
Earthshine in 1999. We thank  Ph. Gastellu-Etchegorry for
providing us Earth observation satellite data. We also thank C.
Prigent, L. Beaufort, G. Bellucci, D. Gillet, S. Le Mou\'elic, A.
Labeyrie, J.M. Perrin and referee S. Franck for constructive
remarks about this work. F. Valbousquet from {\it Optique et
Vision} lent us some parts for the spectrograph.
\end{acknowledgements}




\end{document}